\documentclass[useAMS,usenatbib]{mn2e}

\usepackage{graphicx}
\usepackage{multirow}

\title[Chemical evolution of the Milky Way]{The role of the Galactic bar in the chemical evolution of the Milky Way}
\author[O. Cavichia et al.]{O. Cavichia$^{1}$\thanks{E-mail: cavichia@astro.iag.usp.br}, M. Moll\'a$^{2}$, R. D. D. Costa$^{1}$ and W. J. Maciel$^{1}$\\
$^{1}$IAG, University of S\~ao Paulo, 05508-900, S\~ao Paulo-SP, Brazil\\
$^{2}$CIEMAT, Avda. Complutense 40, 28040- Madrid, Spain}
\begin{document}

\date{Accepted 2013 November 6.  Received 2013 October 17; in original form 2013 June 24}

\pagerange{\pageref{firstpage}--\pageref{lastpage}} \pubyear{2002}

\maketitle

\label{firstpage}

\begin{abstract}
In the absence of an interaction, central bars might be the most effective mechanism for radial motions of gas in barred spiral galaxies, which represent two-thirds of disc galaxies. The dynamical effects induced by bars in the first few kpc of discs might play an important role in the disc profiles in this region. In this work, a chemical evolution model with radial gas flows is proposed in order to mimic the effects of the Milky Way bar in the bulge and inner disc. The model is an update of a chemical evolution model with the inclusion of radial gas flows in the disc and bulge. The exchange of gas between the cylindrical concentric regions that form the Galaxy is modelled considering the flows of gas from and to the adjacent cylindrical regions. The most recent data for the bulge metallicity distribution are reproduced by means of a single and longer bulge collapse time-scale (2 Gyr) than other chemical evolution models predict. The model is able to reproduce the peak in the present star formation rate at 4 kpc and the formation of the molecular gas ring. The model with a bar predicts a flattening of the oxygen radial gradient of the disc. Additionally, models with radial gas flows predict a higher star formation rate during the formation of the bulge. This is in agreement with the most recent observations of the star formation rate at the centre of massive barred spiral galaxies. 
\end{abstract}


\section{Introduction}

Since the seminal papers of \citet{eggen62}, \citet{tinsley78}, \citet{searle78}, and \citet{shaver83} on the Milky Way galaxy (MWG) chemical evolution, many studies were performed both theoretically \citep[e.g.][]{ferrini94, molla95, chiappini97, hou00, chiappini01, ballero07} and observationally \citep[e.g.][]{maciel94, vilchez96, henry99, costa04, zoccali08}. In general, these works, among others, focused on studying independently the chemical evolution of each Galactic region. Except for the exchange of matter through gas infalling from the halo to the disc, these chemical evolution models do not consider interactions among other Galactic structures, as for example the disc and the bulge via radial gas flows. 

The effects of radial gas flows on the Galactic abundance gradients were first analysed by \citet{tinsley78}, who concluded that they cannot be neglected in chemical evolution models. Subsequently, \citet{mayor81} considered radial gas flows in a chemical evolution model in the particular case of inflow driven by the infall of gas with zero angular momentum. Later, \citet{lacey85} proposed a more realistic model, suggesting radial flows driven by three different mechanisms: a) infalling gas with a lower angular momentum than the gas rotating in the disc produces an inflow in the disc at a velocity of a few km/s; b) viscosity in the rotating gas layers might cause inflows in the inner regions and outflows in the outer regions with velocities of the order of 0.1\,km/s; c) interactions of the gas with a bar or spiral density wave in the stellar disc will cause radial inflows and outflows in the regions inside and outside the corotation, respectively, with velocities of $\sim$0.3\,km/s at a distance of 10\,kpc from the Galactic Centre. Posteriorly, \citet{goetz92}, studying theoretically how to form a radial gradient of abundances, found that radial flows alone cannot produce abundance gradients, but they are an efficient process to modify (flattening or steepening) one generated by other processes such as a slow infall. More recently, \citet{spitoni11} developed a chemical model for the Galactic disc, including radial flows as a dynamical consequence of infall of gas in the disc. A variable inflow speed is required in their model to explain the observed gradients. Their radial flows scheme was applied by \citet{mott13} to study the effects of gas flows at high redshifts.  They concluded that radial flows are very important to reproduce the abundance gradients and their temporal evolution. \citet{schonrich09} proposed a model with radial mixing of gas and stars that is able to fit the observational data of the Geneva-Copenhagen survey \citep{nordstrom04}. This model was updated by \citet{bilitewski12}, who proposed a simple analytic scheme to describe the coupling of infalling gas from the intergalactic medium and radial flows within the disc based on angular momentum conservation.

In the absence of interactions, central bars might be the most effective mechanism for radial motions of gas and mixing in barred spiral galaxies, which represent two-thirds of disc galaxies.  \citet{pagel79} first suggested that barred spirals may have flatter gradients compared to normal spirals, a pattern clearly supported in the more extensive work by \citet{martin94}, who found direct relations between the slope of the observed oxygen abundance gradient of a barred spiral and the galaxy's bar strength (ellipticity) and length in the sense that stronger bars are accompanied by flatter gradients. This empirical result is consistent with numerical models  \citep[e.g. ][]{athanassoula92, friedli94, ann05} in which the presence of a bar enhances large-scale mixing over the galaxy's disc, damping radial abundance variations. Recent numerical simulations \citep{minchev12} have predicted that bars can induce large-scale gas flows across the discs of these galaxies with rates in the range of $0.1-10 \mbox{ M}_{\sun} \mbox{yr}^{-1}$, but resulting in a non-linear behaviour, since they produce a flattening in the disc at regions located beyond the outer Lindblad resonance (OLR) and simultaneously they may create a steeper radial distribution in the inner disc. In fact this is also found observationally; for example, \citet{elmegreen09} found an inflow of gas of $\sim 6\, \mbox{M}_{\sun} \mbox{yr}^{-1}$ in the inner disc of NGC~1365 while it is $\sim 40\,\mbox{M}_{\sun} \mbox{yr}^{-1}$ out of the OLR. This implies that the gas accumulates in a ring near the inner Lindblad resonance (ILR), which in turn may provoke an increase in the star formation rate \citep[SFR;][]{zurita04} and, consequently, higher abundances than those expected in the disc.

Another example of the effects of bars is the changes observed in bulges of external spiral galaxies, as can be seen in the work of \citet{coelho11}.  The authors found a bimodal age distribution in the bulges of barred spiral galaxies based on a stellar population synthesis investigation, with mean ages of 4.7 and 10.4 Gyr. Unbarred galaxies do not show this bimodal distribution, perhaps implying that bars have the effect of rejuvenating bulges through gas flows towards the centre of these galaxies driven by the secular evolution of the bars. Similar results were found by \citet{ellison11}, where the study of a sample of 294 galaxies with visually classified, strong, large-scale bars is compared to that of a control sample of unbarred disc galaxies, showing that bars can also enhance the central star formation in massive barred galaxies. Recently, \citet{fisher13} have measured the molecular gas in the centre of spiral galaxies, comparing the results for barred and unbarred ones and relating them with the star formation and the stellar masses. They found that galaxies with high molecular gas densities in their bulges are mainly barred. However, they also found that classical bulges with low gas surface density may also be barred. They suggested that in order to understand the connection between the existence of a bar and the gas surface density in the bulges, it is necessary to take into account both phases of gas.

The dynamical effects induced by the galactic bars in the first few kpc of the galactic discs might play an important role in the disc profiles at this region \citep{gutierrez11, dimatteo13, fisher13}. In fact, \citet{portinari99} already noticed that even using a set of different star formation laws, none was able to reproduce at the same time the metallicity gradient and the radial gas profiles of the Milky Way inner disc. This problem was addressed by \citet[][hereafter PC00]{portinari00}, who developed a chemical model with radial gas flows in order to reproduce, at the same time, the metallicity gradient and the radial gas profile of the disc. Their successful model includes a specific gas velocity pattern that mimics the dynamical effects of the Galactic bar and reproduces the peak in the gas distribution around 4 kpc. 

Regarding the bulge, the recent observational data of \citet{babusiaux10}, \citet{hill11} and \citet{bensby13} point that the Galactic bulge has two main populations. One metal-poor, which reflects the classical bulge component formed on a short time-scale and giving rise to the old spheroid population, and a metal-rich population, including a bar kinematics and probably originated in a longer time-scale driven by the secular evolution of the Galactic bar, forming the pseudo-bulge. Therefore, given these results, the Galactic bulge could have the characteristics of both classical bulge and pseudo-bulge. This scenario was predicted by \citet{wyse92} and \citet{samland03}, the latter by means of a dynamical model, and recently by \citet{tsujimoto12}, \citet{grieco12} and \citet{obreja13}. In the model by \citet{grieco12}, the metal-rich population is explained by assuming a delay relative to the formation of the metal-poor population, formed out of a substantially enriched gas, and on a longer time-scale of $\sim$3 Gyr. On the other hand, in the model by \citet{tsujimoto12}, the metal-rich population is formed from an inflow of gas from the inner thin disc in a time-scale of $\sim$~4~Gyr, although the details of such inflow in the inner disc are not explored in their work. \citet{obreja13} analysed the bulges resulting from cosmological simulations. They showed an important early starburst resulting from the collapse-like fast phase of mass assembly, followed by a second phase with lower star formation, driven by a variety of processes such as disc instabilities and/or mergers. These authors suggested that bulges of massive spiral galaxies will generically have two bulge populations that could be connected to classical and pseudo-bulges. 

The present paper is a further step in the study of the relation between radial gas flows and the chemical evolution of the Milky Way disc and bulge. Our main goal is to analyse in detail the role of radial gas flows not only in the abundance gradient of the disc, but also in the formation and evolution of the Galactic bulge. Moreover, we can fit our models to observational data of the disc that are much richer than were available to PC00. This paper is organized as follows: Section \ref{sec:model} describes the chemical evolution model, the bulge parameters used, and the inclusion of the radial gas flows. Section \ref{sec:model} also describes the Galactic bar parameters and the effects of the bar on the radial gas flows. The description of  the observational data which the models will be compared against is given in Section \ref{sec:obs}. The results of models with radial gas flows applied to the bulge and the disc are shown in Section \ref{sec:results}. This section also presents the results from the model with radial gas flows within the disc based on angular momentum conservation. In Section \ref{sec:conc} the summary and conclusions are presented.

\section{The chemical evolution model}
\label{sec:model}

The chemical evolution model is an update of the model developed by \citet[][their model using the radial mass distribution 28 with efficiencies corresponding to $N=4$, hereafter MD05]{molla05}, where the authors presented a generalization of the multiphase chemical evolution model applied to a wide set of theoretical galaxies with different masses and evolutionary rates. The version of the model presented by MD05 is a generalization of the model developed by \citet{ferrini92} for the solar neighbourhood, and later applied to the whole Milky Way disc by \citet{ferrini94} and other spiral galaxies by \citet{molla96} and \citet{molla99}. The model was also applied to the Milky Way bulge and the bulges of external spiral galaxies in \citet{molla95} and \citet{molla00}. The main differences of the model presented in this work and that presented by MD05 is the inclusion of radial gas flows and the central bulge in order to compute the chemical evolution of the main components of the Galaxy together. This is a more realistic approach, since the interactions via radial flows between the bulge and disc can be taken into account in the present model. In this section we briefly describe some important aspects of the model, referring to the original papers for a more detailed description. 

\subsection{Bulge parameters}

The central bulge was included in the code following \citet{molla95} and \citet{molla00}. In these models, the bulge is an oblate spheroid flattened on the Galactic plane with a semi major axis $x=2$ kpc and a semi minor axis $z=1.5$ kpc \citep{kent92}. 

\begin{figure}
 \begin{center}
   \includegraphics[scale=0.6]{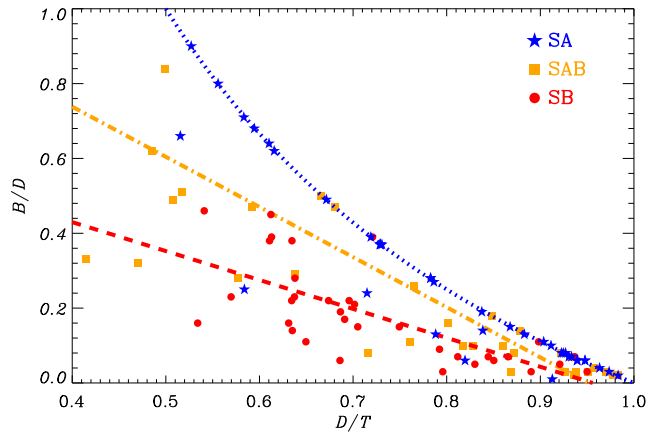}	
   \caption{Bulge-to-disc light ratio ($B/D$) as a function of the disc-to-total light ratio ($D/T$) for three different Hubble types: SA, SAB, and SB. The blue dotted line represents the relation $B/D=(D/T)^{-1}-1$, for the SA galaxies. The orange dash--dotted and red dashed lines denote a linear fit for the SAB and SB galaxies, respectively.\label{fig:bdxdt}}
 \end{center}
\end{figure}

In order to maintain the generalizations of the model introduced by MD05, the bulge mass ($M_{\mbox{\scriptsize b}}$) is now computed from the relation between $M_{\mbox{\scriptsize b}}$ and  $M_{\mbox{\scriptsize d}}$, the disc mass. This relation is obtained from a correlation between the bulge-to-disc light ratio ($B/D$) and the disc-to-total light ratio ($D/T$) for SB galaxies. \citet{weinzirl09} performed two-dimensional bulge--disc--bar decomposition on $H$-band images of 143 bright, high mass ($M_\star \ge 1.0\times10^{10}\,\mbox{M}_{\sun}$), and low-to-moderately inclined ($i < 70^\circ$) spirals. Fig. \ref{fig:bdxdt} shows the bulge-to-disc light ratio as a function of the disc-to-total light ratio for SA, SAB and SB galaxies, as classified by \citet{weinzirl09}. 

The total light is defined by \citet{weinzirl09} as $T=D+B+Bar$. In the case of the SA galaxies, there is no contribution of the bar luminosity, so that $T=D+B$. Therefore, for the SA galaxies, the bulge-to-disc light ratio as a function of the disc-to-total light ratio can be written as $B/D=(D/T)^{-1}-1$. This function is shown as the dotted curve in Fig.~\ref{fig:bdxdt}. The $B/D$ versus $D/T$ relations for the SAB and SB galaxies do not follow this curve, since the bar luminosity is not included in the bulge luminosity. A linear fit to this relation resulted in the dash--dotted and dashed lines for the SAB and SB galaxies, respectively, as shown in this figure. The SAB galaxies have the linear relation

\begin{equation}
B/D = -(1.3\pm0.2)\times D/T + (1.3\pm0.1),
\label{bd}
\end{equation}
with a strong linear correlation coefficient of -0.81. Similarly, the linear fit for the SB galaxies resulted in

\begin{equation}
B/D = -(0.8\pm0.1)\times D/T + (0.7\pm0.1),
\label{bd}
\end{equation}
with a moderate linear correlation coefficient of -0.77. 

Considering the MWG, in our models the disc optical radius is $\sim 13\,\mbox{kpc}$, the disc mass and the total mass at this radius are $8.79\times10^{10}$ and $12.83\times10^{10} \,\mbox{M}_{\sun}$, respectively. Assuming a mass-to-light ratio of 1 \citep{kent92}, then $D/T\sim0.69$ and, from equation (\ref{bd}), $B/D \sim 0.20$. This results in a bulge mass of $1.85\times10^{10}\,\mbox{M}_{\sun}$, in excellent agreement with the value of $1.8\times10^{10}\,\mbox{M}_{\sun}$ obtained by \citet{sofue09}. 

\begin{figure}
 \begin{center}
   \includegraphics[scale=0.6]{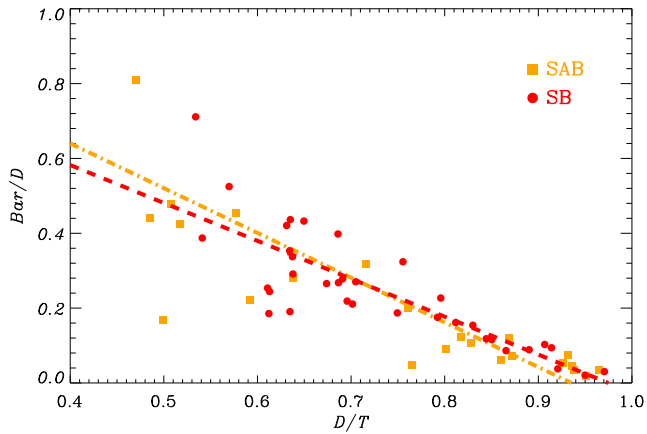}	
   \caption{Bar-to-disc light ratio ($Bar/D$) as a function of the disc-to-total light ratio ($D/T$) for SAB and SB galaxies. \label{fig:bardxdt}}
 \end{center}
\end{figure}

From Fig. \ref{fig:bdxdt} there are some superpositions of SAB and SB galaxies in the plane $B/D$ versus $D/T$, although a linear fit resulted in different linear relations, as shown in the figure. This is explained by Fig. \ref{fig:bardxdt}, where the bar-to-disc light fraction as a function of the disc-to-total light fraction is shown.  The linear fit resulted in a slope and an intercept of $-1.2\pm0.2$ and $1.1\pm0.1$ for SAB galaxies, and $-1.0\pm0.1$ and $0.99\pm0.09$ for SB galaxies, respectively. The linear correlation between $Bar/D$ versus $D/T$ is strong, with a linear correlation coefficient of -0.83 for both types of spiral galaxies. The bar-to-disc light fractions of these galaxies are very similar, so that the $Bar/D$ contribution for the total light is almost the same in the cases of SAB and SB galaxies. This cause the superposition of SAB and SB galaxies in the plane $B/D$ versus $D/T$. It is interesting to note that the $Bar/D$ light ratio can be inferred from the $D/T$ light ratio in Fig.~\ref{fig:bardxdt}. This can be used in the study of the role of bar formations and the 
relation between gas inflows into the central kpc and the secular formation of pseudo-bulges. Thus, quantifying bar properties, such as the fractional light and disc mass ratio ($Bar/D$), can yield insight into these processes.

In \citet{molla95}, the parameters for star and cloud formation in the bulge were calculated by following their radial variation from the disc towards the Galactic Centre. Here we follow \citet{molla00}, where the bulge is a distinct structure from the disc, formed by the infall of gas from the halo. But, instead of treating the bulge as an isolated structure with respect to the disc as in \citet{molla00}, in this work the bulge and the disc are allowed to interact and exchange gas via radial gas flows. This is very common in barred spiral galaxies in which bars are an efficient way to transport gas towards the centre of these galaxies \citep[e.g.][]{friedli94, elmegreen09}. The details of the radial gas flows will be explained in Section \ref{sec:flows}.

\subsection{Element production and chemical evolution}

The Galaxy is assumed as a three-structure system with a halo, a disc, and a bulge. The Galaxy is divided into concentric cylindrical regions, each 2 kpc wide. The total initial halo mass is assumed to be in the form of gas. The total mass $M$ and its radial distribution $M(R)$ are calculated from the respective rotation curves $V(R)$, and these rotation curves are derived from the universal rotation curve from \citet{persic96}. From the rotation curve, it is possible to calculate the radial mass distribution with the expression: 

\begin{equation}
 M(R) = 2.32 \times 10^5 R \ \ V(R)^2.
\end{equation}

The halo gas collapses to fall in the equatorial disc, originating the disc or bulge  as a secondary structure. The infalling gas from the halo is parametrized by $f\times{g_{\mbox{\scriptsize h}}}$, where $g_{\mbox{\scriptsize h}}$ is the halo gas mass and $f$ is the infall rate. The mass which remains in the halo gives a ratio $M_{\mbox{\scriptsize h}}/M_{\mbox{\scriptsize d}}$ for the baryonic component. The collapse time-scale, which defines this infalling gas rate, depends on the galactocentric distance through the expression:

\begin{equation}
 \tau(R) = \tau_c\mbox{e}^{\frac{R-R_{c}}{R_d}},
 \label{eq:tau}
\end{equation}
where it is assumed that the total mass surface density follows the surface brightness exponential shape. In this equation, $\tau_{\mbox{\scriptsize d}}=4$ Gyr, $R_{\mbox{\scriptsize d}} = 4.06$ kpc is the disc-scale length and $R_{\mbox{\scriptsize c}} = R_{\mbox{\scriptsize opt}}/2 = 6.5$ kpc is the characteristic disc radius, where $R_{\mbox{\scriptsize opt}}$ is the disc optical radius. The reader is referred to e.g. MD05 and \citet{ferrini94} for details. Regarding the bulge, in order to reproduce the giant star metallicity distribution function (MDF), the collapse time-scale is fixed to 2 Gyr in this work, which is longer than the disc collapse time-scale projected in the direction of the Galactic Centre. 

In the multiphase model, the star formation is a two-step process: first the diffuse gas forms molecular clouds. Then stars form from cloud--cloud collisions or by the interaction between massive stars and the molecular gas. The former is called spontaneous star formation and the latter induced star formation.
The mass in the different phases of each region changes by the processes associated with the stellar formation and death by the star formation due to spontaneous fragmentation of gas in the halo; formation of gas clouds in the disc from the diffuse gas; induced star formation in the disc due to the interaction between massive stars and gas clouds; and finally, the restitution of the diffuse gas associated with these processes of cloud and star formation. In the halo, the SFR for the diffuse gas follows a Schmidt law with a power index $n = 1.5$. In the disc, the stars form in two steps: first, molecular clouds are formed from the diffuse gas also following a Schmidt law with $n=1.5$ and a proportionality factor $\mu$. The stars are produced by a spontaneous process associated with cloud--cloud collisions at a rate which is proportional to an efficiency $H'$, and a stimulated star formation process proportional to an efficiency $a'$ is assumed.

The enriched material is originated from the stars at the end of their evolution, taking into account their nucleosynthesis, initial mass function (IMF) and the final fate, either through a quiet evolution or by Type I and II supernova explosions. 

The model uses an IMF from \citet{ferrini90} given by
\begin{equation}
\log(\Phi(m)) = 0.3 + \frac{-(0.73 + a \times(1.92+2.07\times a))^{0.5}}{m^{1.52}},
\end{equation}
where $a = \log m$. 

The original model was modified to include metallicity-dependent yields. The yields for high-mass stars are from \citet{woosley95}, while those for low mass stars are from \citet{gavilan05}. Type I supernova releases are considered including the model revised by \citet{iwamoto99}. The SNIa rates are computed using the data provided by \citet{ruiz-lapuente00}, as described in \citet{gavilan06}.

\subsection{Radial flows \label{sec:flows}}

\subsubsection{Model equations}

The complete set of first-order non-linear integrodifferential equations for each zone to be solved is given by \citet{molla96} and \citet{ferrini92}. As mentioned by \citet{ferrini92}, this system of equations may be treated as a purely differential system by choosing a suitable time interval $\Delta \tau$, inside which the integral terms could be considered constant. This time interval is also the timestep between the model outputs, which are stored in the memory during the lifetime of the MWG. We assume $\Delta \tau \approx 10^6 \, \mbox{yr}$ and we were able to check the stability of the solution and, consequently, the goodness of the approximation by varying $\Delta \tau$ by a factor of 2 around this value. Since a star of mass 100~M$_{\odot}$ will die in a time-scale of $3.5\,\mbox{Myr}$, using $\Delta \tau \approx 10^6 \, \mbox{yr}$ we make sure that the time resolution is sufficient to compute the chemical evolution of the most massive star in our model.

Here we just describe the processes associated with the variation of the gas mass and the chemical evolution inside each radial zone. The time variation of the gas mass $g$ is given by

\begin{equation}
\frac{dg}{dt}=-\mu g^n + a'c \ s_2 + H' c^2 + fg_{\rm H} + W + F_{\mbox{\scriptsize rf}}.
\label{eq:dg}
\end{equation}

The first term in this equation is related to the cloud formation from diffuse gas, the second is the remaining gas from the induced star formation via massive star--cloud interaction, the third is the remaining gas after the star formation from cloud collisions, the fourth is the infalling gas from the halo and the fifth is the restitution of the mass due to the death of stars. The novelty in equation (\ref{eq:dg}) is the last term $F_{\mbox{\scriptsize rf}}$, the radial flows between radial zones. This term is defined as

\begin{equation}
F_{\mbox{\scriptsize rf}}(k) = F_2(k) - F_1(k),
\end{equation}	
where $F_1(k)$ and $F_2(k)$ are functions that define the flow of gas that is leaving and entering the shell $r_k$. Each of these functions can be separated into two other functions as 

\begin{figure}
 \begin{center}
   \includegraphics[width=6cm]{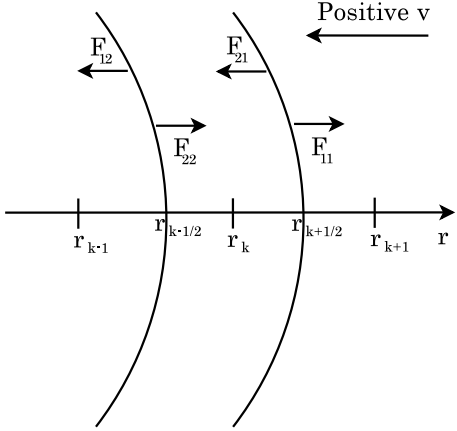}	
   \caption{Scheme for the radial flows in the shell $r_k$.  \label{flows}}
 \end{center}
\end{figure}

\begin{equation}
F_1(k) = F_{11}(k)+F_{12}(k),
\end{equation}

and

\begin{equation}
F_2(k) = F_{21}(k)+F_{22}(k).
\end{equation}	

These function are illustrated in Fig. \ref{flows} and correspond to mass flows in the following directions.

\begin{enumerate}
 \item Inwards the shell $r_k$ from the shells $r_{k+1} \mbox { and } r_{k-1}$, which correspond to $F_{21}$ and $F_{22}$, respectively. 
 \item  Outwards the shell $r_k$ to the shells  $r_{k+1} \mbox{ and } r_{k-1}$, which correspond to $F_{11}$ and $F_{12}$, respectively.
\end{enumerate}

The flows in the $k$ shell can be written in terms of the gas surface density $\sigma(k)$ and the velocity  $v_k$ as

\begin{equation}
F_{11}(k) = -2\upi r_{k+1/2} v_{k+1/2}[\ \chi(-v_{k+1/2})\times \sigma_k], 
\end{equation}
	
\begin{equation}
F_{12}(k) = 2\upi r_{k-1/2} v_{k-1/2}[\ \chi(v_{k-1/2})\times \sigma_k], 
\end{equation}	

\begin{equation}
F_{21}(k) =2\upi r_{k+1/2} v_{k+1/2}[\ \chi(v_{k+1/2})\times \sigma_{k+1}], 
\end{equation}	

\begin{equation}
F_{22}(k) = -2\upi r_{k-1/2} v_{k-1/2}[\ \chi(-v_{k-1/2})\times \sigma_{k-1}].
\end{equation}

Since the equations of our models depend on the gas mass instead of the gas surface density, the latter can be written in terms of the gas mass $g(k)$ as 

\begin{equation}
\sigma_k=\frac{g_k}{2\upi r_{k} \Delta r_k},
\end{equation} 
where $\Delta r_k = r_{k+1/2} - r_{k-1/2}$ is the width of each ring and is fixed to be 2 kpc. The function $\chi$ is a step function defined as:

\begin{equation}
\chi(v_k) = \left\{ \begin{array}{ll}
1 & \textrm{if $v_k>0$}\\
0 & \textrm{if $v_k < 0$}.\\
  \end{array} \right.
\end{equation}

In the case of  $v_k>0$ (flow towards the Galactic Centre), $F_{11}$ and $F_{22}$ vanish, while $F_{12}$ and $F_{21}$ are positive. In this case, $F_{1}=F_{12}$ and $F_{2}=F_{21}$. On the other hand, if $v_k<0$ (flow towards the Galactic anticentre), $F_{12}$ and $F_{21}$ vanish and  $F_{1}=F_{11}$ and $F_{2}=F_{22}$. We assume that there are no flows from the outer regions of the Galactic disc where star formation is not present. This is an acceptable approximation since \citet{bilitewski12} found that radial accretion from the outer disc gas reservoir should be small.

In the innermost shell, we have $k\, = \, 0$ and in this case $v_{-1/2} = 0$, since we cannot have inflows from an inner region. On the other hand, outflows are allowed with the following flow equation:

\begin{equation}
F_{11}(k=0) = v_{1/2}\times \frac{g_0}{\upi(\Delta r_0/2)^2}.
\end{equation}

 In the subsequent shell, the inflow term from the $k=0$ shell can be written:

\begin{equation}
F_{22}(k=1) = -v_{1/2}\times \frac{g_0}{\upi(\Delta r_0/2)^2}.
\end{equation}

The set of equations for the chemical evolution in each ring are obtained from \citet{ferrini92} and are given by:

\begin{equation}
\frac{dX_{i}}{dt} = \frac{-\psi X_{\rm i} + fg_{\rm H} X_{i \rm H}+ W_{i}}{(g+c)} + \Delta_{i}.
\label{eq:xi}
\end{equation}

All terms in this equation depend on the index $k$ and is omitted to alleviate the notation. $X_{i}$ are the abundances by mass of 15 chemical elements: H, D, $\phantom{}^3$He, $\phantom{}^4$He, $\phantom{}^{12}$C, $\phantom{}^{13}$C, $\phantom{}^{16}$O, $\phantom{}^{14}$N, $\phantom{}^{20}$Ne, $\phantom{}^{24}$Mg, $\phantom{}^{28}$Si, $\phantom{}^{32}$S, $\phantom{}^{40}$Ca, $\phantom{}^{56}$Fe, and the neutron-rich isotopes synthesized from $\phantom{}^{12}$C, $\phantom{}^{13}$C, $\phantom{}^{16}$O, and $\phantom{}^{14}$N. $\psi$ is the SFR determined self-consistently. The restitution of the mass due to the death of stars ($W_{i}$) is calculated from

\begin{equation}
 W_{i}(t) =  \int_{m_{0}}^{m_{\rm f}} \! \left[ \sum_j \tilde{Q}_{ij}(m)X_j(t-\tau_m)\right]\Psi(t-\tau_m) \, \mathrm{d}m,
\end{equation}
where $ \tilde{Q}_{ij}(m)$ are the restitution matrices \citep[see e.g.][]{ferrini92}, $m_{0}=0.1$, $m_{\rm f}=100 \mbox{ M}_{\sun}$ and $\tau_m$ is the main-sequence lifetime of a star of mass $m$. The parameter $c$ is the cloud mass in the disc. The last term $\Delta_{i}$ is the increase or decrease of $X_{i}$ due to the flows of gas and is defined as:

\begin{eqnarray}
\Delta_{i}(k)=\frac{X_{i}(k+1)-X_{i}(k)}{g_k+c_k}\times F_{21}(k) \nonumber \\ 
 + \frac{X_{i}(k-1)-X_{i}(k)}{g_k+c_k}\times F_{22}(k). 
\end{eqnarray}

As discussed by \citet{lacey85}, gas flows in the disc can be generated by the viscosity $\nu$ in the gas layer. This viscosity will induce an inward flow in the inner parts of the Galaxy and an outward in the outer parts with velocity $|v| \propto \nu/r$.  The authors estimate that $\nu \approx 1 \mbox{ km s}^{-1} \mbox{ kpc}$ for H\,{\sc i} clouds. In the case of a uniform radial flow pattern $v_{\rm r}$, the velocities of the flows are given by $v_k = v_{\rm r}/r_k$, where $0 < v_{\rm r} < 1 \mbox{ km s}^{-1}$. Another argument on the limit of the flow velocity is the mass conservation. Radial flows must be compensated by the formation of stars in the centre. On the contrary, the disc will deviate from a $1/r$ profile and will become unstable \citep[see][]{struck-marcell91}. Massive star formation will ensure that the excess of mass is expelled out of the region through some sort of wind phenomenon similar to that produced by supernova explosions. Therefore, to maintain a $1/r$ profile, there should exist a limit on the radial flow velocity. This limit depends on the total local surface mass density and on the SFR in the central region \citep{struck-marcell91}. For example, \citet{struck-marcell91} gives the following relation between the SFR and gas surface densities and the flow velocity:

\begin{equation}
 \Sigma_{\mbox{\scriptsize SFR}} \approx 20 \left( \frac{ \Sigma_{\mbox{\scriptsize Gas}} \ v_{\rm r}}{R}\right) \ \frac{\mbox{M}_{\sun}}{\mbox{Gyr} \ {\mbox{pc}^{-2}}}.
 \label{eq:sfr}
\end{equation}

Radial flows in barred galaxies are funnelled to the Galactic Centre along the bar. Hence, we can assume that the flow going to the central kpc of the MWG corresponds to 10 per cent of the total flow of the disc. Using the Kennicutt--Schmidt law with a power index $n=1.4$  and equation (\ref{eq:sfr}), we can find that

\begin{equation}
v_{\rm r} \approx 1.25 \left(\frac{\Sigma_{\mbox{\small Gas}}}{1 \ \mbox{M}_{\odot}  \mbox{pc}^{-2}}\right)^{0.4} \left(\frac{R}{1 \ \mbox{kpc}}\right) \mbox{km s}^{-1}.
\end{equation}

Taking the gas surface density of the radial flow at 3 kpc to be $\sim 5 \mbox{ M}_{\odot} \mbox{pc}^{-2}$,
we can find that a flow of velocity $v_{\rm r} \sim 7 \mbox{ km s}^{-1}$ could sustain the SFR at the MWG central kpc. If the bar induces a radial flow with a velocity larger than this limit, there should be some mass-loss. A stronger flow will contribute to dilute the enriched material ejected from the evolved stars and, consequently, the flattening of the radial abundance gradient is expected in this case.

\subsubsection{The Galactic bar}

The dimensions and orientation of the Galactic bar are not well established. This is the result of our location inside the Galaxy, where any determination of these parameters is difficult because of the reddening by the Galactic dust layer, which obscures the stellar components of the Galaxy in the optical wavelengths. A complete review about the parameters of the bar is out of the scope of this paper and the reader is referred to \citet{dwek95}, \citet{stanek97}, \citet{bissantz02}, \citet{babusiaux05}, \citet{cabrera-lavers07}, \citet{rattenbury07}, \citet{gonzalez11}, \citet{wang12} and references therein. These works, among others, have shown that in the central region of the Galaxy, there is a large-scale bar with a radius of $\sim2.5$ kpc oriented $15\degr -45 \degr$ with respect to the Galactic Centre line of sight. The pattern speed has been estimated from different methods and is somewhat uncertain, with a wide range $\Omega_{\rm b} \sim 35 - 60 \mbox{ km s}^{-1} \mbox{ kpc}^{-1}$ \citep{gerhard11}.

The simulations of \citet{friedli94} pointed that in a stationary barred potential, a net gas inflow is observed inside the bar CR, and a net outflow is observed between the CR and the OLR. The so-called forbidden velocities in the $(\ell,v)$ diagram for $\phantom{}^{12}$CO and the similar results for H\,{\sc i} 21~cm \citep[see e.g.,][]{dame01} are also a direct signature of non-circular orbits in the inner Galaxy, which is interpreted as inner Galaxy gas flows caused by a barred Galactic potential \citep{binney91}. Indeed, \citet{englmaier99} have performed hydrodynamical simulations for the gas dynamics in the inner disc and noticed a substantial mass inflow in this region.  

In order to include the effects of the Galactic bar in the inner few kpc of the MWG, the radial flow velocity pattern described by PC00 in their fig. 15 is adopted. In this case, the Galactic bar ends in correspondence to its CR radius around 3 kpc, and it has an OLR around 4.5 kpc. Gas-dynamical models \citep[e.g.][]{englmaier99, fux99} give a range of $ 3 <  R_{\rm CR} < 4.5$ kpc. In this work, the bar CR radius is $R_{\rm CR} \sim 3$ kpc, as in PC00. In this case, the gas flow is outwards from the CR towards the OLR where it accumulates. The age of the Galactic bar ($T_{\rm bar}$) is not well established and \citet{cole02}, by using data from the Two Micron All-Sky Survey (2MASS), proposed that the MWG bar is likely to have formed more recently than 3 Gyr and must be younger than 6 Gyr. As $T_{\rm bar}$ is not very well constrained, three different ages are considered: 3, 4.5 and 6 Gyr. In this case, the Galactic bar has formed 3, 4.5 and 6 Gyr ago, respectively, and the flows induced by the Galactic bar started at $t_{\rm bar} = 13.2 \mbox{ Gyr} - T_{\rm bar}$. For the sake of simplicity, we will show the results for $T_{\rm bar} = 3$ Gyr and discuss the results if $T_{\rm bar}$ is changed to  4.5 and 6 Gyr.

\begin{figure}
 \begin{center}
\includegraphics[width=8cm]{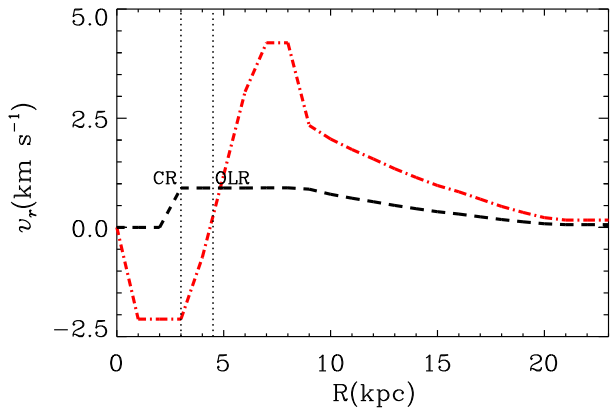}
\caption{Velocity pattern of the model with radial gas flows. The dashed line shows the flow velocity pattern for the model with no bar (model RFC). We defined $t_{\rm bar}$ as $t_{\rm bar} = 13.2 \mbox{ Gyr} - T_{\rm bar}$. At $t \ge t_{\rm bar}$, the Galactic bar starts to contribute to the gas flows at the inner Galaxy, and the flow velocity pattern is represented by the dash--dotted line in the case of model RFB (see Table \ref{tab_models}). Three different values for $T_{\rm bar}$ are considered: 3, 4.5 and 6 Gyr (see the text). The vertical dotted lines mark the positions of the CR and the OLR radii. \label{flow_pat}}
 \end{center}
\end{figure}

 On the other hand, inflows in the disc are expected to occur during the entire evolution of the Galaxy \citep[PC00,][]{lacey85, spitoni11,bilitewski12}.  We have computed two variable flow velocities that are shown in Fig. \ref{flow_pat}. The dashed line is the flow velocity pattern for the case where there is no bar. In this case, there is an inflow of gas in the direction of the Galactic Centre. In this model, it is considered that the disc radial flows start at the beginning of the MWG formation and last to the present time. The dash--dotted line in the same figure represents the model with the bar-like gas flows. In this case, for $t < t_{\rm bar}$ there are no gas flows in the Galaxy, and the chemical evolution is the same as in the case of MD05. As the time increases, after a time $t \ge t_{\rm bar}$, the Galactic bar starts to change the radial flow pattern in the inner disc. At this moment, the flows begin with the velocity pattern described by a dash--dotted line in Fig. \ref{flow_pat}. The plus and minus signs in the velocities give the direction of the flows. As already mentioned, a positive velocity means that the flow is towards the centre of the Galaxy, while a negative velocity means that the flow is in the opposite direction. In order to reproduce the data available for the inner Galaxy (bulge and disc), two independent scalefactors are used: $\xi_1$ and $\xi_2$. These parameters are multiplicative factors of the velocity at a given position $k$. The final velocity is given by

\begin{equation}
v_k = \left\{ \begin{array}{ll}
\xi_1\times v_{\rm r}(k), & \textrm{if $R \le 8 \mbox{ kpc}$}\\
\xi_2 \times v_{\rm r}(k), & \textrm{if $R > 8 \mbox{ kpc}$},\\
  \end{array} \right.
\end{equation}
where $v_{\rm r}(k)$ is the original velocity proposed by PC00. We varied the values of the parameters $\xi_1$ and $\xi_2$ between 0 and 20 in steps of 0.1. Here we just show the main results obtained by varying the parameters  $\xi_1$ and $\xi_2$. We should stress that in our models the radial inflows are faster than the radial outflows, as can be seen in Fig. \ref{flow_pat}. This causes a net gas flow towards the centre of the Galaxy. Therefore, we should expect an increase of mass of gas in the central regions of the MWG and, consequently, an increase of the nuclear SFR. This will be explored in more detail in Section \ref{sec:results_bulge}.

\subsubsection{Angular momentum conservation in the disc \label{angmom}}

Infalling gas from the halo with a lower angular momentum than the gas rotating in the disc produces an inflow in the disc at a velocity of a few $\mbox{km s}^{-1}$ \citep{lacey85}. Recently, \citet[][hereafter BS12]{bilitewski12} proposed a simple analytic scheme to describe the coupling of infalling gas from the intergalactic medium and radial flows within the disc based on angular momentum conservation. In their model, only the case of infalling gas with a lower angular momentum than the disc gas was considered. Thus, the gas always flows radially inwards and each ring just receives mass from its outer neighbour. BS12 give the following relation between the radial flow velocity $v_{\rm r}$ and the difference of velocities of the infalling gas $v_{\rm i}$ and the gas rotating in the disc $V_{\rm c}$: 

\begin{equation}
v_r(R,t) = -\frac{f\times g_{\rm H}}{g_{\rm D}}\frac{(v_{\rm i} -V_{\rm c})R}{V_{\rm c}},
\end{equation}
where $f\times g_{\rm H}$ is the rate of infalling gas in the disc from the halo, and $g_{\rm D}$ is the disc gas mass in the correspondenting ring. The velocity of the infalling gas is parametrized by BS12 assuming a linear relationship with the galactocentric distance:

\begin{equation}
v_{\rm i}(R)=V_{\rm c}\left(b+a\frac{R}{R_{\mbox{\scriptsize out}}}\right).
\end{equation}
In this equation, $b$ and $a$ are free parameters and $R_{\mbox{\scriptsize out}}$ is the disc outer radius. By changing the parameters $a$ and $b$ and comparing the model results with the Fe/H radial gradient from Cepheid stars, BS12 were able to find the best --fitting parameters. In their fig. 3, they show the resulting flow velocities for some models at four different times: 3, 6, 9 and 12 Gyr. Thus, we can test the flow velocities proposed by BS12 in our models with radial gas flows and to compare the results from the flows originated due to the angular momentum conservation in the disc with the flows caused by the Galactic bar.

\section{Observational data}

\label{sec:obs}

The data sets which the models will be compared to are described in this section.

\subsection{Bulge sample}

Data for the bulge red giant branch (RGB) stars are taken from \citet{johnson13}. They presented chemical abundance ratios of [Fe/H], [O/Fe], [Si/Fe] and [Ca/Fe] for 264 RGB stars in three Galactic bulge fields located at $(\ell,b)=(-5.5,-7)$, $(-4,-9)$ and $(+8.5,+9)$. Their results are based on equivalent width and spectrum synthesis analyses of medium-resolution ($R\sim18\,000$), high signal-to-noise ratio (S/N$\sim75-300 \mbox{ pixel}^{-1}$) spectra obtained with the Hydra spectrographs on the Blanco 4m and WIYN 3.5m telescopes. \citet{hill11} analysed a sample of 219 red clump stars in Baade's window from $R=20\,000$ resolution spectra obtained with FLAMES/GIRAFFE at the Very Large Telescope (VLT). \citet{alves-brito10} and \citet{bensby13} provide abundance ratios of [O/Fe], [Mg/Fe] and [Si/Fe] for bulge stars. The latter also provide abundance ratio of [Ca/Fe]. Their observations are obtained with high-resolution ($R = 45\,000 \mbox{ or } 67\,000$), moderate signal-to-noise ratio (S/N$\sim 45-100$) optical spectra of 25 bulge giants in Baade's window with the H\,{\sc i}RES spectrograph on the Keck-I telescope. Similarly, \citet{bensby13} obtained high-resolution ($R \sim 45\,000 \mbox{ and } 29\,000$) spectra of 58 bulge dwarf stars with the ESO Very Large Telescope and H\,{\sc i}RES spectrograph on the Keck-I telescope, respectively, that were obtained while the stars were optically magnified during gravitational microlensing events. \citet{ryde10} provide abundance ratios of [O/Fe] and [Si/Fe] for 11 bulge giants, observed at high resolution, in the $H$ band using the CRIRES spectrometer mounted on the
ESO Very Large Telescope. \citet{rich12} report abundance analysis for 30 bulge giants based on $R=25\,000$, infrared spectroscopy using the NIRSPEC at the Keck-II telescope.

The chemical abundances of bulge intermediate-mass population is represented by the sample of planetary nebulae (PNe) from \citet{cavichia10} and \citet{escudero04}, hereafter IAG-USP sample. This sample comprises low-resolution ($R\sim$1000) and moderate signal-to-noise ratio spectra of 100 PNe obtained with the 1.60m telescope of the National Laboratory for Astrophysics
(LNA, Brazil). 

\subsection{Disc sample \label{sec:discsample}}

Abundance gradients in the Milky Way disc are commonly traced by measuring the abundance of metals such as O, Ne, S, Ar, and Fe as a function of the Galactocentric distance. The first four elements are mostly observed in emission line objects like H\,{\sc ii} regions and PNe, while Fe is mostly observed in the absorption lines of stars. \citet{esteban05} performed echelle spectrophotometry of eight Galactic H\,{\sc ii} regions with the Very Large Telescope Ultraviolet Echelle Spectrograph. \citet{balser11} measured radio recombination line and continuum emission in 81 disc H\,{\sc ii} regions with the National Radio Astronomy Observatory Green Bank Telescope. The data of six H\,{\sc ii} regions obtained from \citet{rudolph06} were observed in the far-IR using the Kuiper Airborne Observatory and combined with Very Large Array radio continuum observations to determine the abundances of O, N and S. The authors combined these data with reanalyzed literature data, resulting in a data set of 117 H\,{\sc ii} regions. The disc data for PNe are obtained from \citet{henry10}, where spectrophotometric observations of 41 Galactic PNe are presented, which were combined with the data of \citet{henry04} and \citet{milingo10}, providing a completely homogeneous sample of 124 Galactic PNe. This sample was recently classified by \citet{maciel13} in groups of young and old objects according to their individual ages. Their best method for age determination was adopted, resulting in a final PN sample of young objects with ages of the order of 1 Gyr.

The Fe radial gradient is measured from the Cepheid variable stars' data of \citeauthor{andrievsky02a} (2002a,b) and \citet{luck11}. The first two studies provide Fe abundance determination for 101 Galactic Cepheids based on high-resolution ($R\sim80\,000$), high signal-to-noise ratio (S/N$ > 100$) spectra obtained at the Anglo-Australian Observatory. \citet{luck11} give Fe abundances for  226 Cepheids obtained from high signal-to-noise (S/N$\sim$100) and high-resolution (R$\sim$30\,000) spectroscopic observations with the Hobby--Eberly Telescope.

The disc radial profile of the diffuse gas surface density is obtained from \citet{nakanishi03}, who derived a three-dimensional distribution of H\,{\sc i} gas in the MWG using the latest H\,{\sc i} survey data cubes and rotation curves. The similar data for the molecular gas surface density profile are taken from \citet{williams97} and \citet{olling01}. The former fitted the mass distribution of molecular clouds using cloud catalogues of two CO surveys of the first quadrant. The latter presented a tabulated version of the radial variation of the molecular hydrogen calculated from axisymmetric Milky Way mass models. The SFR radial profile is compared with the data of \citet{stahler05}, which compiled data of observations from supernova remnants, pulsars and H\,{\sc ii} regions. 

\section{Results}
 \label{sec:results}

We ran several models by varying the parameters of the velocity pattern of the flows, $\xi_1$ and $\xi_2$, in the cases with and without the bar. The parameters of the best models are summarized in Table \ref{tab_models}.  The first column in this table indicates the name adopted for the model and the second the gas flow velocity pattern. The third column shows the velocity $v_{\rm r}$ in the case of the constant velocity models and the fourth and fifth columns give the scalefactors used in the pattern velocity of PC00. We now turn to the results of the models.

\begin{table}
  \begin{center}
  \caption{Model parameters used.}
  \label{tab_models}
  \begin{tabular}{l c c c c c}\hline \hline
Model & Pattern & $v (\mbox{km s}^{-1}) $& $\xi_1$ & $\xi_2$ \\ 
\hline
MD05         & Static    & --    & -- &  --  \\ 
RFC           & Inflow (no bar)    & Fig. \ref{flow_pat}  & 2.5 &  2.5\\ 
RFB           & Variable flow (bar) & Fig. \ref{flow_pat} & 15.5 & 8.0 \\
 \hline						     
  \end{tabular}
 \end{center}
\end{table}

\subsection{Galactic bulge}
\label{sec:results_bulge}

\begin{figure}
 \begin{center}
\includegraphics[width=8cm]{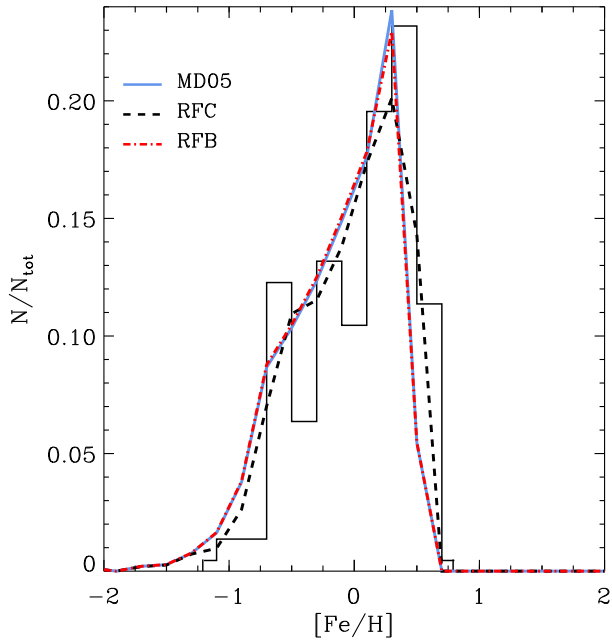}
\caption{Iron distribution (histogram) for 219 red clump stars from \citet{hill11} compared with our models using different radial flows. The continuous blue line shows the reference model of MD05 without radial gas flows. The dashed black line shows the model with a radial inflow velocity given by Fig. \ref{flow_pat} (model RFC). The dash--dotted red line shows the model with radial flows including the effects of the bar (model RFB). Iron abundances are displayed in the notation [Fe/H]$ = \log(\mbox{Fe/H})-\log(\mbox{Fe/H})_{\sun}$. \label{mdf}}
 \end{center}
\end{figure}

Fig. \ref{mdf} shows the MDF for the bulge giant stars normalized for the total number. The histogram corresponds to the data from \citet{hill11}. The model predictions are represented by lines and are given in terms of the solar reference value from \citet{asplund09}. Our models give the time evolution of the total SFR. To obtain the theoretical number of stars in the giant phase, we used the Padova isochrones revised by \citet{molla09}, as well as the IMF of \citet{ferrini90}, in the range of mass from 0.8 to 8 M$_{\sun}$.  The reference model of MD05 without radial gas flows is shown in the figure with the continuous line. The observational data of \citet{hill11} are better fitted by our models with a collapse time-scale of $\tau=2$ Gyr. Including a constant inflow of gas (model RFC) the bulge MDF moves to higher metallicities, although the peak of the distribution remains at the same position of [Fe/H]~=~0.1~dex. The main difference with respect to the reference model is at the value of the peak of the distribution, decreasing the number of stars with [Fe/H]~=~0.1~dex in the model with radial inflow. Model RFB with the bar does not change dramatically the bulge MDF. It is important to note that models with radial flows do not show differences in the MDF in a level that observations would be able to constrain, given the actual observational data uncertainties. 

\begin{figure}
 \begin{center}
\includegraphics[width=8cm]{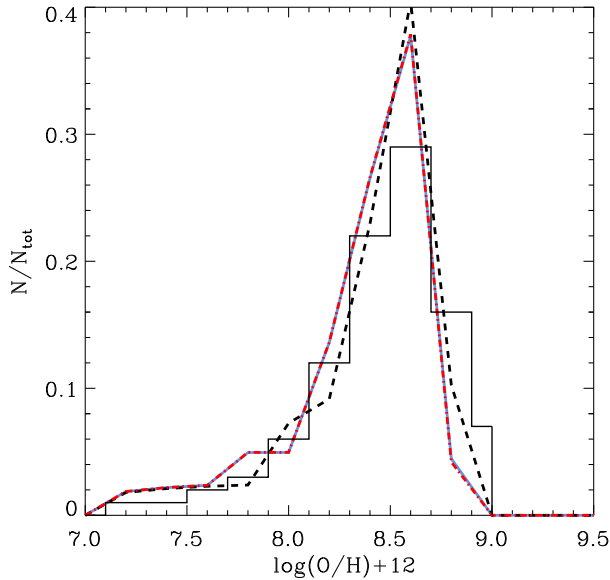}
\caption{O/H distribution (histogram) for 100 PNe from \citet{cavichia10} and \citet{escudero04} (IAG-USP sample) compared with our models using different radial flows. The oxygen abundances are displayed as $\log(\mbox{O/H})+12$. Line styles are as in Fig. \ref{mdf}. \label{oh_hist}}
 \end{center}
\end{figure}

Oxygen abundances in PNe reflect the interestellar abundances at the time the progenitor star was born, although a small depletion may be observed due to ON cycling for the more massive progenitors. The bulge O/H abundance distribution is shown in Fig.~\ref{oh_hist}. The observational data are 100 PN abundances from the IAG-USP sample. All models are able to reproduce the data,  in spite of the fact that model RFC can fit better the wings of the distribution. The largest discrepancies between the models and observed distribution occur at the peak of the distribution, in which case the models predict a peak higher than the data. This can be due to oxygen depletion in the intermediate-mass population, or due to the incompleteness of the observational sample.

Usually, in order to reproduce the bulge MDF, a scenario is required where the bulge is formed by the fast collapse of primordial gas accumulating in the centre of the MWG, as noted by many authors \citep[][and references therein]{cescutti11}. Nonetheless, we have noticed that we need in our models a collapse time-scale of $\tau=2$ Gyr, which is longer than previous works, to reproduce better the recent asymmetrical bulge MDF of \citet{hill11}. Additionally, a \citet{ferrini90} IMF is used in our models instead of a top-heavy IMF like that of \citet{ballero07}. Frequently, the asymmetrical bulge MDF of \citet{hill11} is interpreted as an MDF composed of two main populations: one metal poor formed with a shorter time-scale and other metal rich formed with a longer time-scale \citep{grieco12, tsujimoto12}. This is not the case in our models, where one single time-scale and a steeper IMF are adopted to reproduce the recent observational data. 

\begin{figure}
 \begin{center}
   \includegraphics[width=8cm]{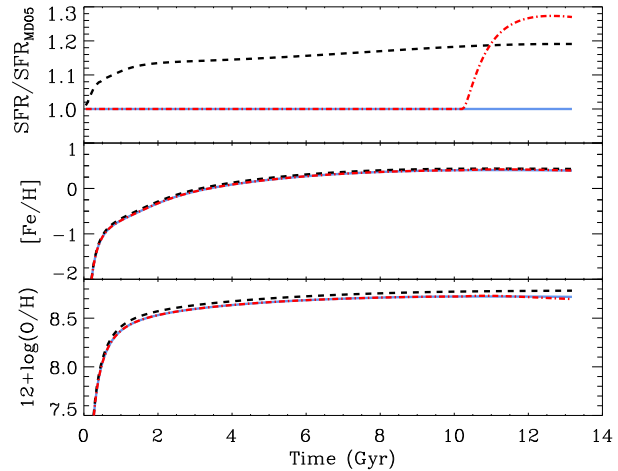}
   \caption{Top: bulge SFR normalized for the bulge SFR of the reference model (MD05) as a function of time. Middle: bulge Fe/H abundance as a function of time. Bottom: bulge O/H abundance as a function of time. The lines styles are as in Fig. \ref{mdf}. \label{sfrfe_time}}
 \end{center}
\end{figure}

Fig. \ref{sfrfe_time} (top) shows the time evolution of the bulge SFR normalized with the bulge SFR of the reference model (SFR$_{\mbox{\scriptsize MD05}}$). The SFR in the bulge increases very fast in the first Gyr of the bulge formation, reaching the maximum value at $t = 0.7$~Gyr. In the case of model RFC, the SFR is higher during the entire bulge evolution. This is so because the inflow of gas in this model increases the gas density in the inner regions of the Galaxy since the bulge formation. In this model, at the end of the bulge evolution, the SFR is increased by 19 per cent compared with the reference model. In the cases of model RFB, the gas flows start at 10.2 Gyr when the bar is formed, and the bulge ratio SFR is increased by 27 per cent compared with the reference model. Since the bulk of the bulge stars form at the first Gyr of the MWG formation, the effects of the radial flows of these models are not seen in the SFR at this point. Only after $t=t_{\rm bar}$, it is possible to note an appreciable difference in the SFR due to radial flows in the cases of the model with bar-like gas flow. It is important to note that in model RFB, the radial inflows are faster than the radial outflows, as can be seen in Fig. \ref{flow_pat}. This causes a net gas flow towards the centre of the Galaxy. Therefore, we should expect an increase of mass of gas in the central regions of the MWG and, consequently, an increase of the nuclear SFR.

The model predictions for the bulge metallicity time evolution are shown in Fig. \ref{sfrfe_time} (middle panel). The [Fe/H] time evolution is very similar for all models. It increases very fast at the beginning of the bulge formation, since the SFR is increasing very fast at this time. After 6 Gyr, it stays almost constant as time increases. Model RFC presents slightly higher [Fe/H] values compared with the other models, since the effects of the gas inflows in the bulge are more important in this model. The same occurs for the O/H time evolution (Fig. \ref{sfrfe_time}, bottom panel).

\begin{figure}
 \begin{center}
 \includegraphics[width=9cm]{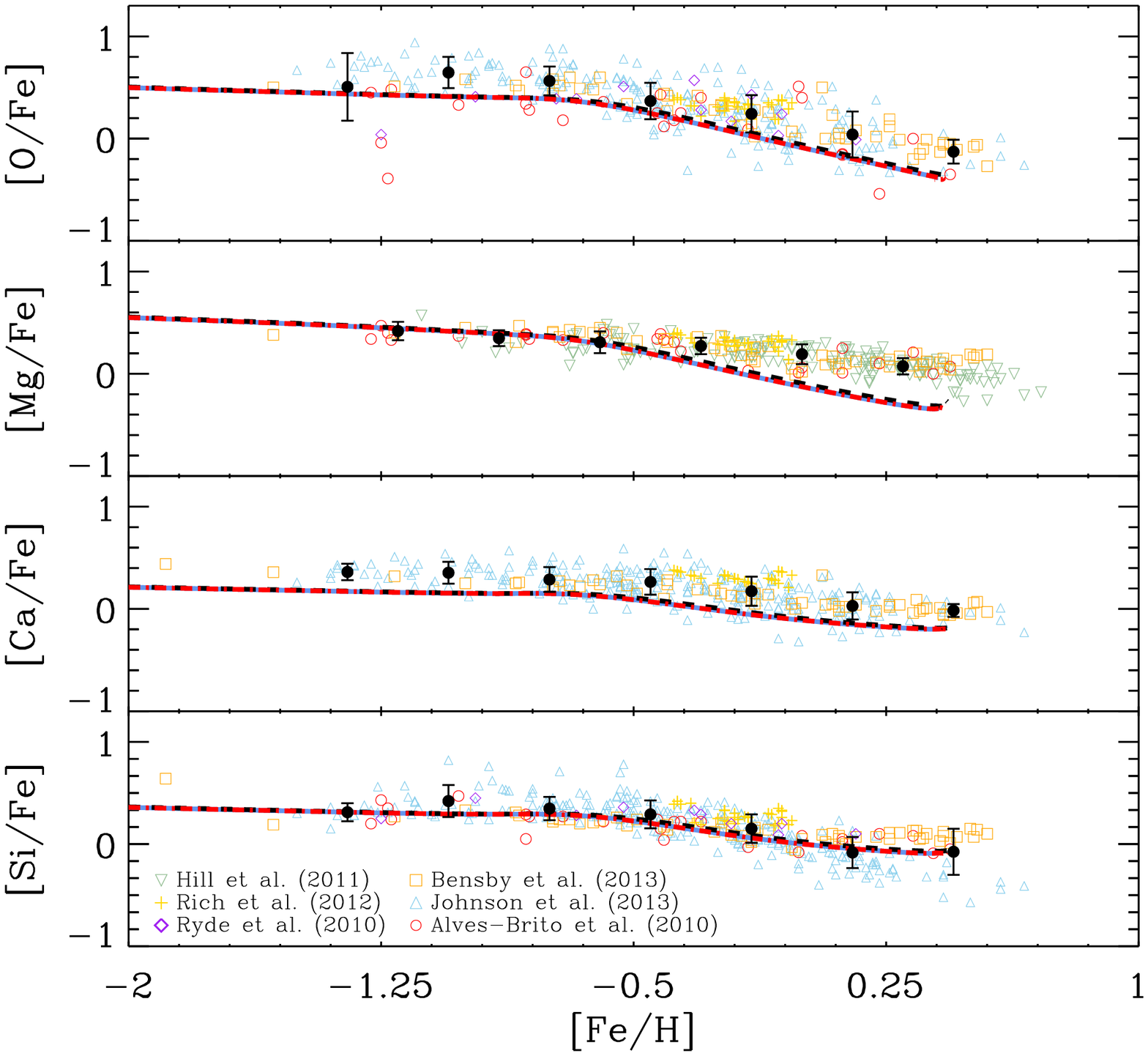}
 \caption{Comparison between the predictions of our model using different flow velocities and the literature data for $[\alpha/\mbox{Fe}]$  where $\alpha$ stands for O, Mg, Ca and Si,  versus [Fe/H]. The observational data for the bulge correspond to open circles from \citet{alves-brito10}, open squares from \citet{bensby13}, open upside triangles from \citet{johnson13}, open upside-down triangles from \citet{hill11}, plus signs from \citet{rich12} and open diamonds from \citet{ryde10}. The filled black circles with error bars denote the mean and standard deviation of the data with [Fe/H] bins of 0.3 dex. The line codes of the models are as in Fig. \ref{mdf}. \label{alpha_ratio}}
 \end{center}
\end{figure}

Our models are also able to predict the evolution of abundance ratios. This is shown in Fig. \ref{alpha_ratio}, where the abundance ratios between O, Mg, Ca, Si and Fe are plotted as a function of [Fe/H]. Our models are able to reproduce the observational data of \citet{alves-brito10} and \citet{bensby13} regarding the [O/Fe] and [Si/Fe] abundance ratios. The [Si/Fe] abundance ratios of \citet{johnson13} are also correctly predicted by our models, but they fail in reproducing their [O/Fe] and [Ca/Fe] data, predicting lower abundance ratios. The [Ca/Fe] abundance ratios of \citet{bensby13} and \citet{rich12} are higher than the results of our models. \citet{johnson13} compared their data with the model of \citet{kobayashi11} and the better agreement is reached with a flat IMF ($x=0.3$) and both outflow and a larger binary fraction than the solar neighbourhood. However, their model still predicts lower [O/Fe] than the observational data for [Fe/H] $ < -0.5$. In our case, it seems that the yields of oxygen for high-mass stars from \citet{woosley95} need to be increased in order to reproduce the most recent data of the bulge. Mg is also underestimated by our models and we manually shifted the model predictions by +0.35 dex in order to reproduce the observational data. In \citet{molla00} this fact was already noted and the underestimation of Mg abundances by our models is probably related with the SN II nucleosynthesis model of \citet{woosley95}. As discussed by \citet{molla00}, to reproduce the Mg abundance, the SN II nucleosynthesis should be two times higher than predicted by \citet{woosley95}. The data show that [Mg/Fe] is less sensible with the Fe production by the SN Ia explosions. Compared with the data,  the models predict a steeper [Mg/Fe] ratio for higher metallicity regime. It seems that in order to the models reproduce the data, Mg production should be increased in both SNe Ia and SNe II.  The stellar yields of Mg from low- to intermediate-mass stars as proposed by \citet{karakas10} should also be considered in order to increase Mg abundance in our models. Since the results of the present paper are not affected by Mg abundances, this will be addressed in a future paper.

We ran several models with different values for $T_{\rm bar}$ and the main conclusions of this section remain unaltered. Changing $T_{\rm bar}$ from 10.2 Gyr to the extreme value of 7.2 Gyr (bar age of 6 Gyr) has little effects on the results of model RFB. Considering $T_{\rm bar} = 7.2$ Gyr, the main differences found are in the bulge MDF. The number of stars in the peak of the MDF decreases while the number of stars with higher metallicities increases. Nonetheless, the peak of the distribution remains at the same position, still reproducing the observed MDF. It is interesting to note that in our models, a young bar of 3 Gyr is able to reproduce the main observational data available today. However, within these simplified models it is not possible to discuss in detail the scenarios for bar formation, its structure and age, and the related gas flows, which is beyond the goals of this paper.

\subsection{Galactic disc}

\begin{figure}
 \begin{center}
  \includegraphics[width=9cm]{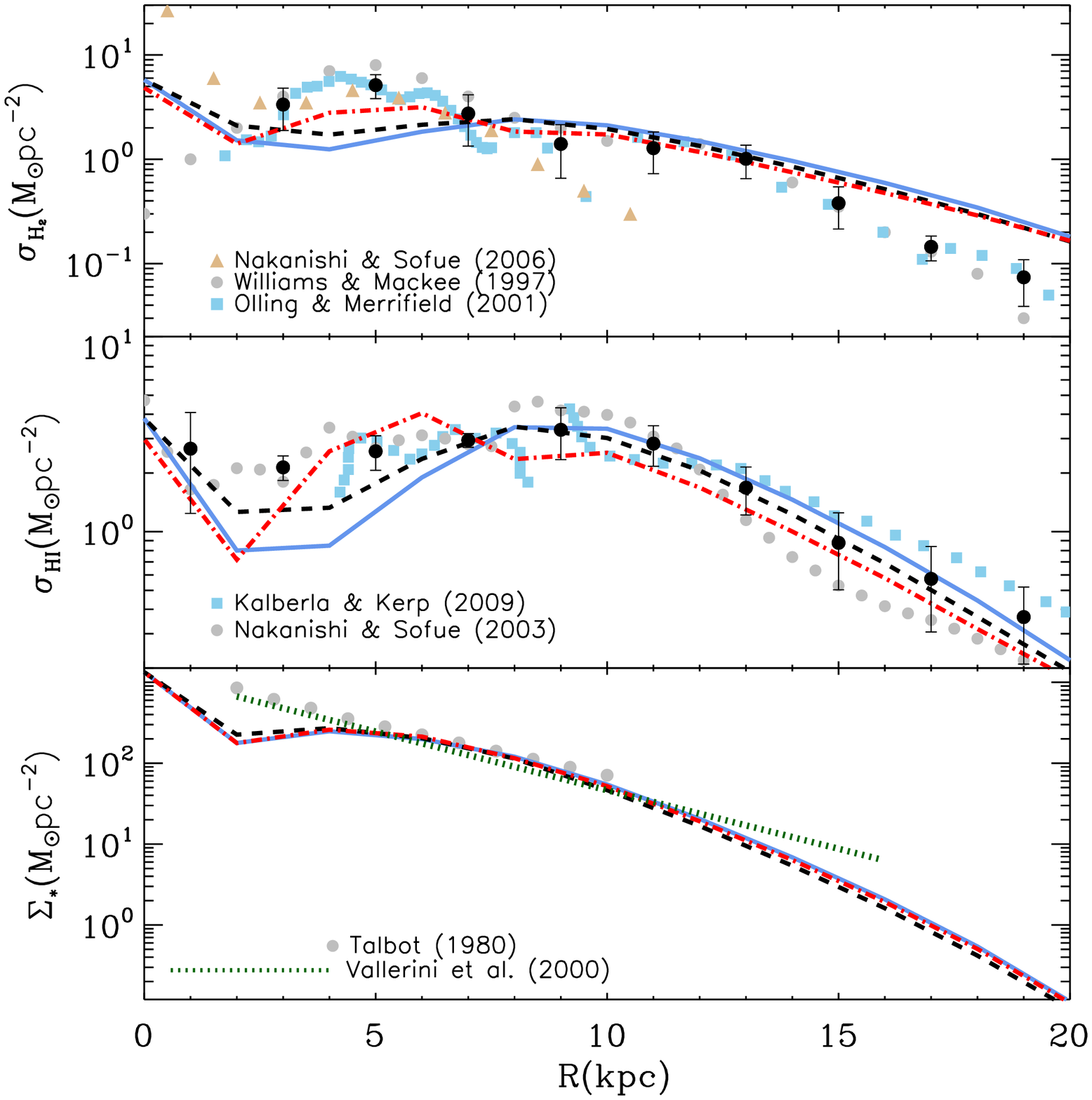}
  \caption{Top: molecular gas surface density radial distribution for each model, compared to the observational data from \citet{williams97} and \citet{olling01}. Middle: diffuse gas surface density radial distribution. The observational data (circles) are from \citet{nakanishi03}. Bottom: stellar surface density with the data from \citet{talbot80}. The lines for the models are as in Fig. \ref{mdf}. The black filled circles with error bars denote the mean and the standard deviation of the data computed in bins of 2 kpc.\label{gasprof}}
 \end{center}
\end{figure}

The surface gas density profile predictions are shown in Fig. \ref{gasprof}. \citet{portinari99} noticed that the gas profile peak at 4 kpc is due to the dynamical influence of the Galactic bar and that static chemical evolution models are not able to reproduce it if other constraints like the disc metallicity gradient and surface gas density profile are fitted as well. The model of MD05 is not an exception in this case. Our reference model predicts a much lower $\sigma_{\mbox{\scriptsize H\,{\sc i}}}$ and $\sigma_{\mbox{\scriptsize H}_2}$ than the observations do for $R < 8$ kpc. The inner disc gas density profiles are better reproduced using radial gas flows, as can be seen in Fig. \ref{gasprof}. Model RFC shows that just an inflow of gas is not sufficient to increase the gas density in the inner disc in a level that matches the observations. In particular, model RFB, which includes the bar, is able to fit the $\sigma_{\mbox{\scriptsize H\,{\sc i}}}$ profile at the inner disc much better than our other models do. This is in agreement with PC00, where only their model with the effects of the bar is able to reproduce the radial gas profile properly. On the other hand, we cannot compare our results in this region with those from \citet{spitoni11}, since their models are focused in the outer disc and effects of the central bar are not considered.

\begin{figure}
 \begin{center}
 \includegraphics[scale=0.45]{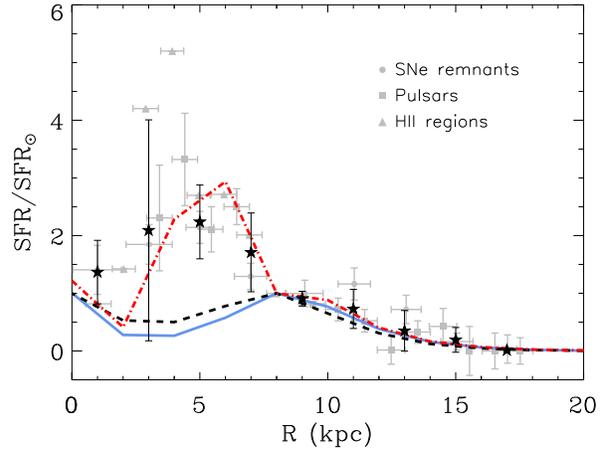}
 \caption{Present SFR as a function of the galactocentric distance. The SFR is normalized for the solar SFR. The observational data are from \citet{stahler05}. The circles represent the SN remnants data, squares the pulsar data and triangles the H\,{\sc ii} region data. The stars with error bars denote the mean weighted by the individual errors in the SFR and the standard deviation of the mean, respectively. Line styles for our models are as in Fig.~\ref{mdf}. \label{sfr_dist}}
 \end{center}
\end{figure}

One advantage of the models with radial flows is that it is possible to perform a reasonable fit to the present disc SFR profile.  The predictions of the models are shown in Fig. \ref{sfr_dist}. The observational data are from \citet{stahler05} and are based on supernova remnants, pulsars and H\,{\sc ii} regions. At the inner disc the data are widely spread in SFR, mainly due to the H\,{\sc ii} regions data from \citet{guesten82}. These data contrast to the other data from SNe remnants and pulsars at the inner disc, showing higher SFR values at $R \sim 3 - 4$ kpc. As already pointed out by \citet{kennicutt12}, the SFR data from H\,{\sc ii} regions need new observations. We followed \citet{kennicutt12} and attributed an error of 0.3 dex to these data. To better visualize the data, we divided the SFR into bins 2 kpc wide and, in each bin, we computed the mean weighted by the individual errors. The stars with error bars in Fig. \ref{sfr_dist} correspond to these averaged SFR. In this figure, we can see that the best fit is achieved with model RFB when the velocity scalefactors $\xi_1$ and $\xi_2$ are 15.5 and 8.0, respectively. That is, in order to reproduce the SFR profile, our models need a faster velocity profile than PC00 used in their work. We can conclude that the present SFR in the MWG inner disc is highly affected by the gas dynamics at this region. Our models indicate that the Galactic bar may play an important role in the SFR of the inner disc. Since we are not performing detailed gas-dynamical simulations of bar formation and the evolution of the gas flows at the inner disc, it is not possible to point out if this is the major effect of the enhancement of the SFR at the inner disc. Considering that our reference model underestimates the SFR at the inner disc, radial flows induced by the bar may be the answer why traditional chemical evolution models are not able to reproduce the SFR at this region.

\begin{figure}
 \begin{center}
\includegraphics[scale=0.45]{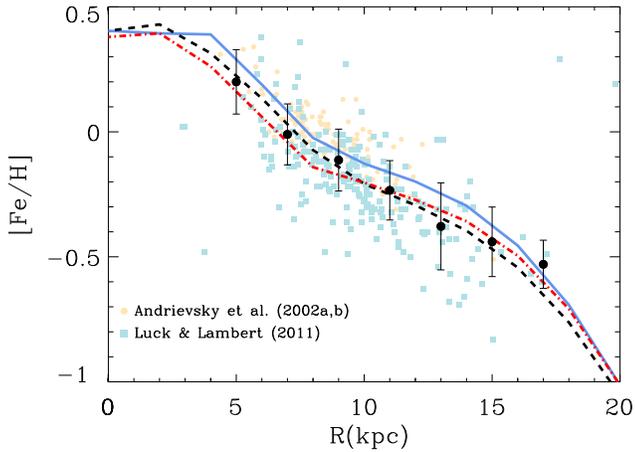}
\caption{Iron radial distribution for Cepheid stars from \citeauthor{andrievsky02a} (2002a,b) (circles) and \citet{luck11} (squares), compared with our models using different radial flows. The black circles with error bars denote the mean and the standard deviation of the data, respectively, in data bins of 2 kpc. Line codes for the models are as in Fig. \ref{mdf}.\label{feh_ceph}}
 \end{center}
\end{figure}

Another important constraint for chemical evolution models is the disc [Fe/H] radial gradient. Fig. \ref{feh_ceph} shows the disc [Fe/H] radial gradient for Cepheid stars from \citeauthor{andrievsky02b} (2002a,b) and \citet{luck11}. All models are in agreement with the data, although the reference model seems to predict a higher gradient level than the recent data of \citet{luck11} indicate. Models with radial flows are better in agreement with the data. The lack of data for the outer region of the disc ($R > 15$ kpc) makes impossible to constrain the models at this region. Our models show that iron abundances continue to diminish as the radial distances increase. More data are needed in the outer disc for a complete view of the iron gradient in this region.

The [Fe/H] radial gradient in the case of  $T_{\rm bar} = 6$ Gyr is steeper in the inner disc ($R < 8$ kpc) and flatter in the outer disc than the case with  $T_{\rm bar} = 3$ Gyr. But the differences are small and given the present uncertainties in the chemical abundances it is not possible to constrain the age of the bar with the models.

\begin{figure}
 \begin{center}
\includegraphics[scale=0.45]{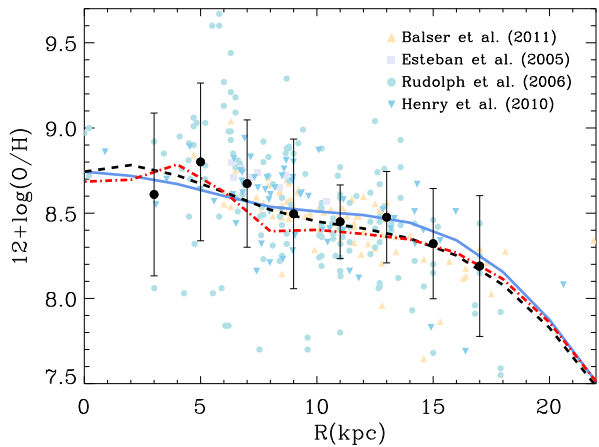}
\caption{Oxygen radial distribution for disc H\,{\sc ii} region from \citet{balser11} (triangles), \citet{esteban05} (squares) and \citet{rudolph06} (circles), compared with our models using different radial flows. The black circles with error bars represent the mean and the standard deviation of the data using a bin size of 2 kpc. The oxygen abundances are displayed as $ \log(\mbox{O/H})+12$. Line styles are as in Fig. \ref{mdf}. \label{oh_hii}}
 \end{center}
\end{figure}

Fig. \ref{oh_hii} shows the oxygen radial gradient from H\,{\sc ii} regions of \citet{balser11} (triangles), \citet{esteban05} (squares) and \citet{rudolph06} (circles). Since the data are widely spread in abundance, to better understand its trend, we divided the data into bins as functions of the galactocentric distance. In each bin, we computed the mean value and the standard deviations. The bin sizes are chosen to be in agreement with the size of the shells of our models. In Fig. \ref{oh_hii}, black circles with error bars denote the mean and the standard deviation of the data with bins 2 kpc wide. The reference model of MD05 is in good agreement with the observational data in Fig. \ref{oh_hii}. However, both models RFB and RFC with radial flows seem to reproduce better the data trend represented by the circles with error bars. The lack of data for the inner disc makes it difficult to compare with our models in this region. It is worth noting that models with radial flows predict  the decrease of the abundances towards the Galactic Centre. This is in agreement with the results of \citet{cavichia11} and \citet{gutenkunst08} from PNe. Comparing Figs \ref{feh_ceph} and \ref{oh_hii}, oxygen shows a flatter gradient when compared with iron, since the time-scales involved in the production of these two elements are different.

It is interesting to note that in Figs. \ref{feh_ceph} and \ref{oh_hii} the inclusion of the gas flows in model RFB steepens the gradient at the inner disc and flattens the gradient at the outer disc. It is well known that barred galaxies display systematically shallower gradients than ordinary spirals \citep[e.g.][]{alloin81, vila-costas92, martin94}. This is likely a consequence of the radial mixing induced by bars \citep{friedli94}. The radial gas flows transport fresh disc material from the outer regions towards the inner ones. Consequently, the density of gas is increased at the inner disc and bulge, and the star formation and the production of metals are also increased. 

In fact, \citet{martin94} found a correlation for external galaxies between the strength of a bar and the metallicity gradient. 
Using their parametrization and the Galactic bar axial ratio of 0.6 from \citet{merrifield04}, we get an O/H radial gradient of $-0.051 \mbox{ dex kpc}^{-1}$, which is compatible with the most recent O/H radial gradient of $(-0.058\pm0.006) \mbox{ dex kpc}^{-1}$ obtained  by \citet{henry10} from PN data. 

\begin{figure}
 \begin{center}
\includegraphics[width=8cm]{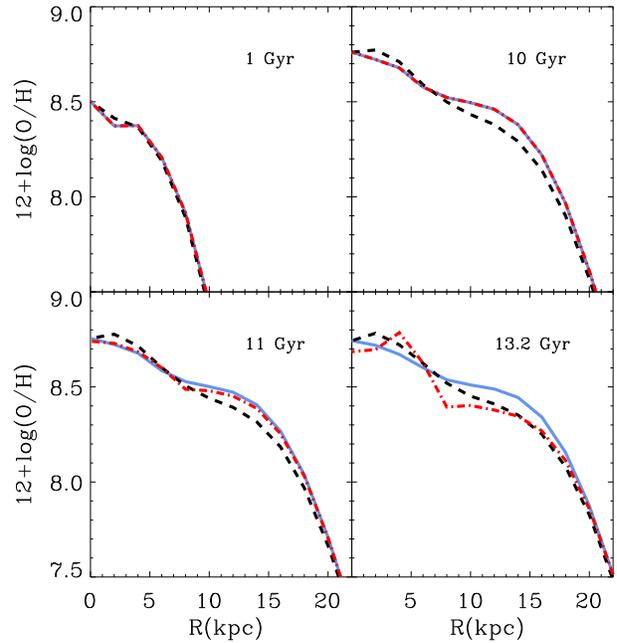}
 \caption{Time evolution of the oxygen radial gradient. At the top right of each figure is indicated the age of the Galaxy at the moment of the simulation. The colours and line styles are as in Fig. \ref{mdf}. \label{grad_oh}}
 \end{center}
\end{figure}

One concern about the evolution of spiral galaxies is the time evolution of abundance gradients.  An interesting discussion about the shape and the temporal evolution of the MWG radial gradient is given by \citet{henry10}. On the one hand, \citet{stanghellini10} conclude that the MWG disc abundance gradient has steepened with time. On the other hand, \citet{maciel09} claim that the gradient has flattened over time. Therefore, the present data ended up to a conflict where both scenarios of flattening and steepening of the gradient are observed. A recent work by \citet{maciel13} based on individual age determinations for the PN central stars can shed some light on this problem. This work suggests that the average oxygen gradient has remained constant during the last 4--5~Gyr.  Resolving this problem about the temporal evolution of the abundance gradient is extremely important from the standpoint of chemical evolution of the MWG. The temporal evolution for the O/H radial gradient for our models is shown in Fig. \ref{grad_oh} for four different ages: 1, 10, 11 and 13.2 Gyr. Clearly, in the models with the bar-like radial flows the gradients are {\it flattening} in the last 3 Gyr, if considered Galactocentric distances are in the range of $8 -15$ kpc. Although the models with the bar-like radial flows predict a flattening of the oxygen gradient, the differences are small compared with the data dispersion in this region. In the outer parts of the disc, for $R > 15$ kpc, the models predict a declining of oxygen abundances as $R$ is increased. To confirm these results, it is needed to extend the observational data to radial distances beyond 20 kpc from the Galactic Centre. Then it will be possible to establish the gradient in terms of its actual value in addition to its shape at the outer disc.

\subsubsection{Angular momentum conservation}

We adopted the radial flow velocity profiles of BS12 to test the effects of the radial flows in the disc due to angular momentum exchange. The reader is referred to the original paper of BS12 for details on the implementation of the model. A brief discussion is given in Section \ref{angmom} of the present paper. 

Two different flow velocity patterns were adopted from the profiles of BS12, corresponding to their model parameters $a=0.5$ and $b=0.3$ (model BS12b03) and $a=0.5$ and $b=0.5$ (model BS12b05) at four different times: 3, 6, 9 and 12 Gyr. In this case, the flow velocity for model BS12b03 increases with the galactocentric distance, while for model BS12b05 it increases up to 11 kpc, where it starts to decrease, as can be seen in their fig. 3. The BS12b05 model has $v_{\rm r} < 0.9 \ \mbox{km s}^{-1}$, while the BS12b03 model has $v_{\rm r} < 3 \ \mbox{km s}^{-1}$. In both models the flow velocities decrease with time.

\begin{figure*}
 \begin{center}
 \includegraphics[width=8.25cm]{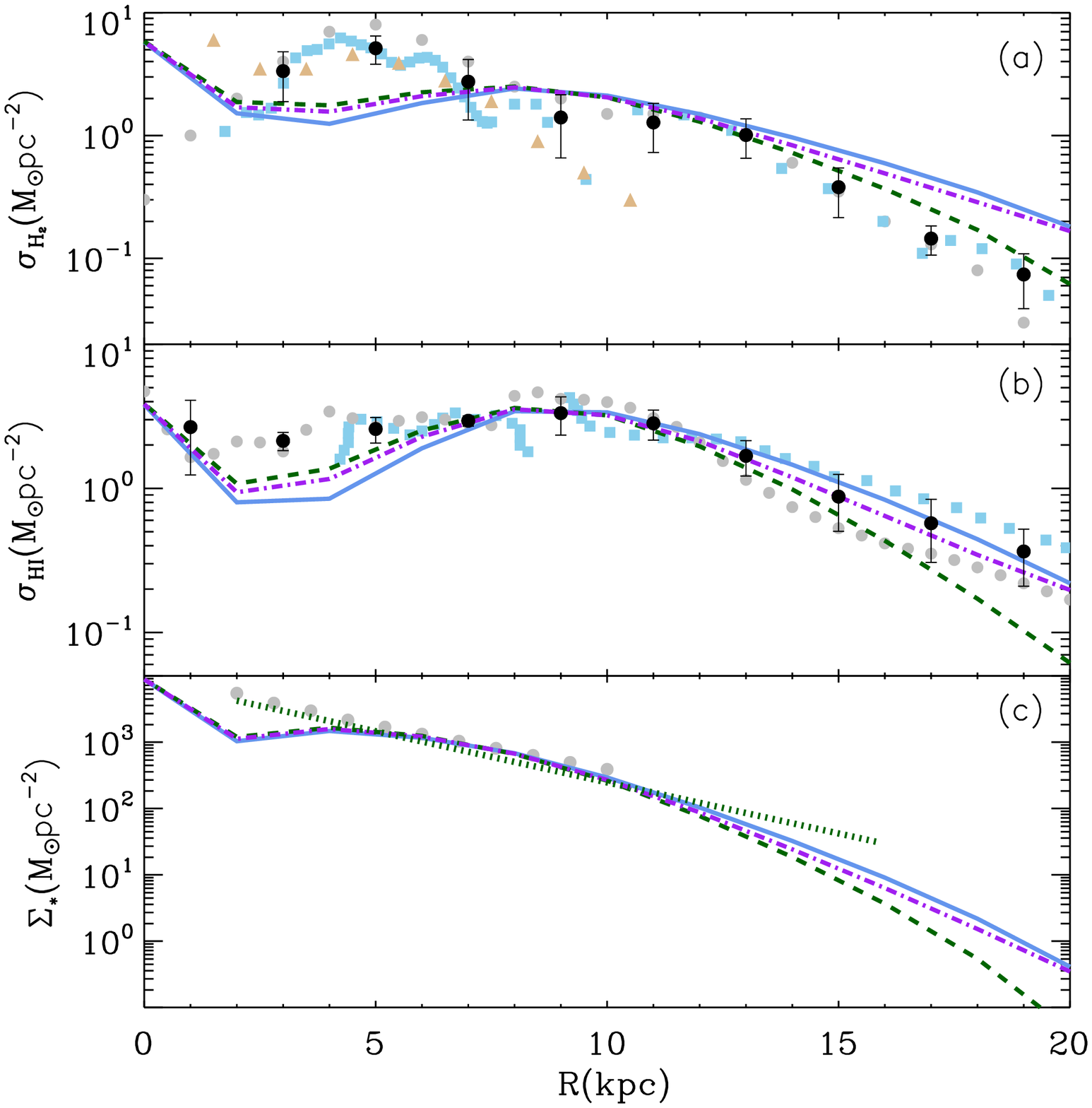}
\includegraphics[width=8cm]{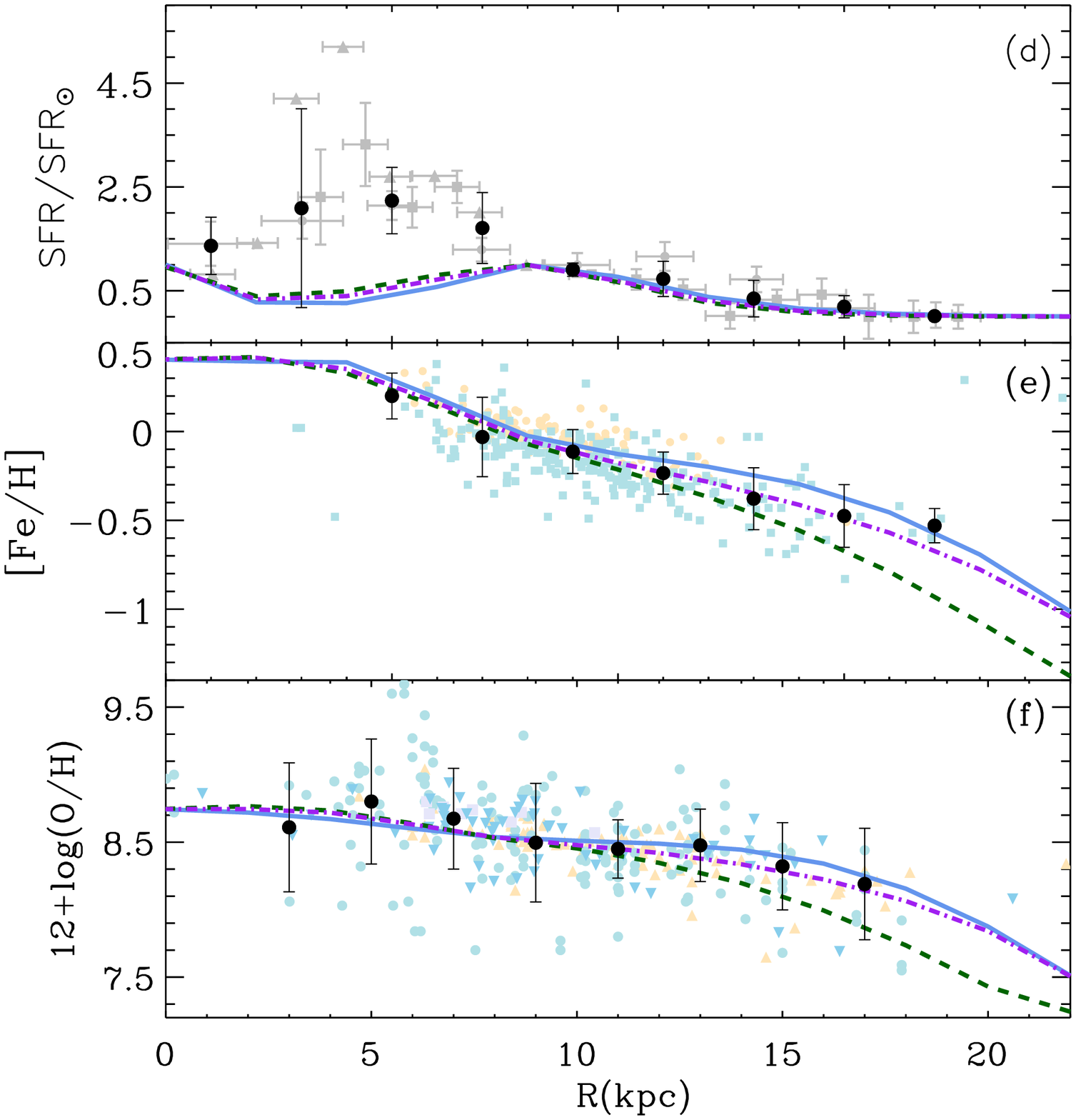}
\caption{Radial distributions for different models with radial gas flows from BS12 with their model parameters $a=0.5$ and $b=0.3$ (model BS12b03, dashed lines), and   $a=0.5$ and $b=0.5$ (model BS12b05, dash--dotted lines) Continuous lines denote the reference model of MD05. (a) Molecular gas surface density. (b) Atomic gas surface density. (c) Stellar mass surface density. (d) Present SFR normalized to the solar value. (e) Fe/H radial abundance gradient. (f) O/H radial abundance gradient. The black filled circles with error bars denote the mean and standard deviation of the data for bins 2 kpc wide. See Section \ref{sec:discsample} for data references. \label{flow_bs12}}
 \end{center}
\end{figure*}

Fig. \ref{flow_bs12} shows the results from models BS12b03 (dashed lines) and BS12b05 (dash--dotted lines) compared with our reference model MD05, continuous line. The BS12b05 model predicts correctly the radial distributions of O/H, Fe/H and stellar mass surface density. But, as in the case of our reference model, it fails in reproducing the SFR, $\sigma_{\mbox{\scriptsize H\,{\sc i}}}$ and $\sigma_{{\rm H}_2}$ in the inner regions of the galactic disc ($R < 8$ kpc). This was expected, since BS12 do not take into account the effects of the galactic bar at the inner regions of the galactic disc. 

The BS12b03 model predicts steeper O/H and Fe/H abundance gradients than the data show.  At the inner disc, the abundance gradients are correctly predicted. Although higher velocities are adopted for this model, the problem of radial distributions of the SFR, $\sigma_{\mbox{\scriptsize H\,{\sc i}}}$ and $\sigma_{{\rm H}_2}$ at the inner disc is still not solved. 

We conclude that, although the BS12 radial gas flows are not able to fit the radial distributions of gas and SFR at $R < 8$ kpc, they can predict correctly the disc abundance gradients of O/H and Fe/H. Therefore, radial gas flows in the disc based on the angular momentum conservation of the infalling gas are important to shape the disc abundance gradients in the sense that they become steeper with the increasing of the flow velocities.

\section{Summary and conclusions}
\label{sec:conc}

In this paper, we have computed a set of chemical evolution models for the MWG bulge and disc where we have considered a scenario involving radial gas flows between concentric radial zones. We have adopted the model of MD05 as our reference model and compared different gas flow pattern velocities with this model. In particular, we investigated the role of the Galactic bar in the chemical evolution of the MWG. It is well known that, in the absence of interactions, central bars might be the most effective mechanism for radial motions of gas and mixing in barred spiral galaxies. Therefore, it is expected that the dynamical effects induced by the Galactic bar in the first few kpc of the Galactic discs might play an important role in the disc profiles in this region \citep{gutierrez11, dimatteo13, fisher13}. In order to investigate these effects on the chemical evolution of the MWG, we have adopted the radial flow scheme of PC00 and our main goal is to analyzse in detail the role of radial flows not only in the abundance gradient of the disc, but also in the formation and evolution of the Galactic bulge. Three different ages for the Galactic bar are considered: 3, 4.5 and 6 Gyr. The main results of this paper can be summarized as follows.

\begin{enumerate}

\item All our models are able to reproduce the bulge metallicity distribution of 219 red clump stars from \citet{hill11}. We considered a scenario where the bulge is formed with a collapse time-scale of 2 Gyr, which is longer than other chemical evolution models predict. Also, a flatter IMF of \citet{ferrini90} is used, rather than a top-heavy IMF of Salpeter or \citet{ballero07}. There is no need to invoke a second infall episode, as recently suggested by other authors \citep{grieco12, tsujimoto12} to explain the observed metallicity distribution.

\item The model with a constant radial inflow of gas and velocities lower than $1 \mbox{ km s}^{-1}$, occurring during the whole MWG evolution, increases the present bulge SFR in 19 per cent, compared with the reference model. A higher star formation in the bulge in this model moves the bulge MDF to higher abundances, although not in a level that might be noted in the observations. On the other hand, models with the bar-like radial flows do not change the bulge MDF, since the bulk of bulge stars are formed in a much shorter time-scale, before the bar formation. The model with the bar-like radial flows (RFB) predicts the highest present bulge SFR, which is 27 per cent higher than the SFR of the reference model of MD05. This is fully in agreement with the results of \citet{ellison11}, showing that bars can also enhance the central star formation in massive barred galaxies.

\item We are able to reproduce the present SFR radial profile at the inner regions of the Galactic disc. In particular, our model with the bar-like radial flows can perform a reasonable fit to the peak of SFR at $\sim4$ kpc, given the actual SFR uncertainties at the inner disc. The present H\,{\sc i} surface density radial profile is also correctly predicted at the inner disc, whereas the H$_2$ radial profile is underestimated at the inner disc, although this model is our best model to predict the H$_2$ radial profile. In general, static chemical evolution models are not able to reproduce this peak if other constraints like the disc metallicity gradient and surface gas density profile are fitted as well.  Our models indicate that the Galactic bar may play an important role in the SFR and gas surface density at the inner disc. 

\item Compared with the reference model, radial flows induced by the bar can flatten the oxygen abundance gradient for $R > 8$ kpc in a relatively short time-scale of 3 Gyr. Additionally, the oxygen gradient is steeper for this model for $4 < R < 8$ kpc. This model with the bar-like radial flows predicts an positive oxygen abundance radial gradient for $R < 4$ kpc, which is in agreement with the observations of PNe in this region \citep{gutenkunst08, cavichia11}. The iron abundance radial gradient is less affected by the radial gas flows, although some differences were found. The same trend of the oxygen radial gradient was noticed with the bar-like radial flows. However, with the present uncertainties in the observations, it is not possible to point which model gives the best fit to the data. The radial flows produce a net accumulation of mass in the inner regions of the disc and may enhance a central star formation. This results in a flatter abundance gradient for the outer disc and in a steeper gradient for the inner disc.

\item A model with radial gas flows in the disc due to angular momentum exchange between the accreted infalling gas and the gas rotating in the disc was proposed by BS12. We tested the effects of these flows in our models and they can correctly predict the disc abundance gradients of O/H and Fe/H, although the gas surface density and the SFR at the inner disc are underestimated.

\end{enumerate}

\section*{Acknowledgments}
   
We acknowledge the anonymous referee for very helpful comments. This work has been financially supported by the grants $\#$2007/07704-2, $\#$2010/18835-3, $\#$2012/01017-1 and $\#$2012/22236-3, S\~ao Paulo Research Foundation (FAPESP), CAPES (Proc. BEX 3026/10-8) and CNPq. O.C. would like to acknowledge M. V. C. Duarte and C. E. Paladini for their help with computational tools, and CIEMAT for their hospitality
during parts of this project. This work has made use of the computing facilities of the Laboratory of Astroinformatics (IAG/USP, NAT/Unicsul), whose purchase was made possible by the Brazilian agency FAPESP (grant 2009/54006-4) and the INCT-A. This work has also been financially supported by the Spanish Government projects ESTALLIDOS:  AYA 2007-67965-C03-02, and Consolider-Ingenio 2010 programme grant CSD2006-00070: First Science with the GTC (http://www.iac.es/consolider-ingenio-gtc). Also partial support  by  the Comunidad de Madrid under grants CAM S2009/ESP-1496 (AstroMadrid) is acknowledged.

\bibliographystyle{mn2e}

\bibliography{biblio}

\end{document}